\documentclass{aa}
\newcommand{\bit}{\begin{itemize}}
\newcommand{\eit}{\end{itemize}}
\newcommand{\ben}{\begin{enumerate}}
\newcommand{\een}{\end{enumerate}}
\newcommand{\bde}{\begin{description}}
\newcommand{\ede}{\end{description}}

\usepackage{graphicx}
\usepackage{lscape}

\usepackage{natbib}
\bibpunct{(}{)}{;}{a}{}{,}

\begin{document}
\title{Internal kinematics of isolated modelled disk galaxies}
\author{W. Kapferer$^1$,
        T. Kronberger$^{1,2}$,
        S. Schindler$^1$,
        A. B\"ohm$^2$, and
        B. L. Ziegler$^2$}

\institute{ $^1$Institut f\"ur Astrophysik,
            Leopold-Franzens-Universit\"at Innsbruck,
            Technikerstr. 25,
            A-6020 Innsbruck, Austria\\
            $^2$Institut f\"ur Astrophysik,
            Universit\"at G\"ottingen,
            Friedrich-Hund-Platz 1,
            D-37077 G\"ottingen, Germany}

\offprints{W. Kapferer, \email{wolfgang.e.kapferer@uibk.ac.at}}

\date{-/-}

\abstract{We present a systematic investigation of rotation curves
(RCs) of fully hydrodynamically simulated galaxies, including
cooling, star formation with associated feedback and galactic
winds. Applying two commonly used fitting formulae to characterize
the RCs, we investigate systematic effects on the shape of RCs
both by observational constraints and internal properties of the
galaxies. We mainly focus on effects that occur in measurements of
intermediate and high redshift galaxies. We find that RC
parameters are affected by the observational setup, like slit
misalignment or the spatial resolution and also depend on the
evolution of a galaxy. Therefore, a direct comparison of
quantities derived from measured RCs with predictions of
semi-analytic models is difficult. The virial velocity
V$_{\rm{c}}$, which is usually calculated and used by
semi-analytic models can differ significantly from fit parameters
like V$_{\rm{max}}$ or V$_{\rm{opt}}$ inferred from RCs. We find
that V$_{\rm{c}}$ is usually lower than typical characteristic
velocities derived from RCs. V$_{\rm{max}}$ alone is in general
not a robust estimator for the virial mass.

 }
\authorrunning {W. Kapferer et al.}
\titlerunning {Internal kinematics of isolated disk galaxies}
\maketitle

%

\section{Introduction}
Spatially resolved rotation curves (RCs) are a fundamental tool to
study the internal kinematics and the distribution of mass in
spiral galaxies. The first discrepancy between the differential
kepler-type of rotation and the rotation of galaxies was detected
by Babcock (1939). These first indirect measurements of
non-luminous matter introduced the concept of dark matter (DM)
into astrophysics. An overview is given by Sofue and Rubin
(2001) and references therein.\\
An important application of RCs lies within a correlation of the
luminosity and the maximum rotational velocity of spirals found by
Tully \& Fisher (1977). The physical origin of the slope and the
scatter of the TFR is still subject to debate. Different
theoretical approaches exist, which differ mainly in the
predictions of the redshift evolution of the TFR. Therefore,
Ziegler et al. (2002) and B\"ohm et al. (2004) used a sample of
field galaxies in the FORS Deep Field to study the TFR at
intermediate redshift. They find a significant change of slope in
comparison to local samples, mainly caused by small, star forming
distant galaxies. However, the measurement of rotational
velocities is more complicated in the case of distant, apparently
small
spirals. \\
In this work we investigate how parameters from models, describing
the shape of RCs, are influenced by observational constraints and
by internal properties of galaxies. Only techniques that take
these systematics into account, as e.g. the method presented in
B\"ohm et al. (2004), can get
robust results for V$_{\rm{max}}$. \\
In recent years, fully N-body/hydrodynamic simulations of spiral
galaxies became an important tool to understand the formation and
evolution of spiral galaxies (e.g. Mihos \& Hernquist 1994;
Springel \& Hernquist 2002). In these simulations cooling, stellar
feedback and galactic winds are taken into account to model
galaxies in a physically motivated way. Here, we extract RCs from
model galaxies simulated with GADGET2 (Springel 2005). We
investigate for intermediate and high redshift galaxies the
influence of large relative slit widths, inclinations and slit
misalignments on the determination of fitting parameters like
V$_{\rm{opt}}$ or V$_{\rm{max}}$. These parameters are commonly
used (e.g. Courteau 1997, Yegerova \& Salucci 2004) as a measure
for the 'peak' circular velocity, e.g. to determine Tully-Fisher
relations.

\section{Simulations}
Recently Kapferer et al. (2005) studied the influence of
galaxy-interactions on the strength and evolution of the star
formation rate of the interacting system. In this work we
investigate the isolated model galaxies, presented in Kapferer et
al. (2005). The initial conditions (hereafter ICs) of the model
galaxies were built according to Springel et al. (2004), which is
based on the work of Mo et al. (1998). The two model galaxies in
this work are chosen such that they represent a Milky Way type and
a small spiral galaxy without any bulge component. In Table
\ref{galaxyproperties} the properties of the ICs of the model
galaxies are listed. The combined N-body/SPH simulation calculates
then 5 Gyr of isolated evolution. For every time step we know the
velocity of each particle and can hence extract realistic rotation
curves.

\begin{table}
\caption[]{Properties of the initial conditions of the model
galaxies}
\begin{tabular}{l l l l }
\hline \hline Properties & Galaxy A & Galaxy B  \cr \hline
circular velocity $V_{\rm{c}}$$^{1}$& 160 & 80 \cr disk mass
fraction$^{2}$ & 0.05 & 0.05  \cr gas content in the disk$^{3}$ &
0.25 & 0.25  \cr disk thickness$^{4}$ & 0.02 & 0.02  \cr  total
mass [$M_{\sun}$] & 1.33x10$^{12}$ $h^{-1}$ & 1.67x10$^{11}$
$h^{-1}$ \cr disk scale length [kpc] & 4.51$h^{-1}$ & 2.25$h^{-1}$
\cr \hline
\end{tabular}
\label{galaxyproperties} $^{1}$... circular velocity at
$\rm{r}_{200}$ in km/s\newline $^{2}$... fraction of disk
particles (stars/gas) in units of halo mass \newline $^{3}$...
relative content of gas in the disk
\newline $^{4}$... thickness of the disk in units of radial scale
length
\newline
\end{table}

\subsection{Rotation Curve Extraction}

\begin{figure}
\begin{center}
{\includegraphics[width=\columnwidth]{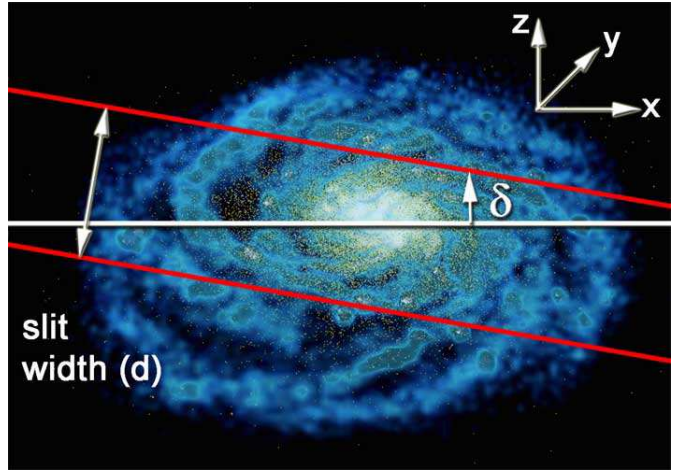}} \caption{Image
of model galaxy A and a virtual slit for extracting a rotation
curve. The slit width d and the slit misalignment angle $\delta$
are indicated.} \label{RCextraktor1}
\end{center}
\end{figure}

\begin{figure}
\begin{center}
{\includegraphics[width=\columnwidth]{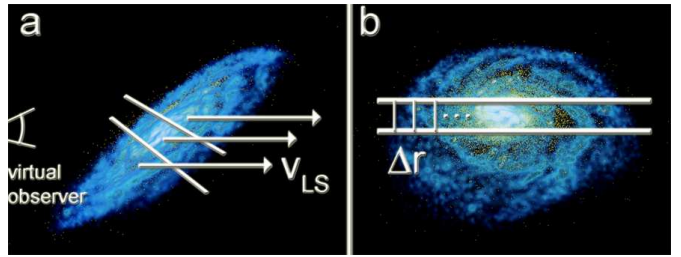}}
\caption{Sketch of our procedure to extract rotation curves of the
model galaxies. (a) the line-of-sight velocity field of a model
galaxy is indicated as it would be observed by a virtual observer.
(b) the sampling along the slit is highlighted ($\Delta$r).}
\label{RCextraktor2}
\end{center}
\end{figure}

In order to extract the rotation curves (RCs) of our simulated
model galaxies, we define a slit with a width d, see Fig.
\ref{RCextraktor1}. In addition we allow for a misalignment angle
$\delta$ to simulate rotations of the slit with respect to the
major axis of the system. Such rotations sometimes occur in
observations using multi-object spectroscopy. In Fig.
\ref{RCextraktor1} the different parameters for the slit are
shown. The slit width d and the misalignment $\delta$ of the slit
with respect to the major axis define the virtual slit. We extract
the RCs from the velocity field of the gas in the following way.
The velocity as a function of radius is determined by averaging
over all line-of-sight velocities of gas particles in thin bins,
with a bin size $\Delta$r along the slit and a side length d
perpendicular to the spatial axis. In Fig. \ref{RCextraktor2} the
velocity field taken for the RC extraction is shown (a). In Fig.
(b) the spatial sampling along the slit is sketched. An RC with a
resolution of 0.1 kpc was extracted and used as a reference.

\begin{figure}
\begin{center}
{\includegraphics[width=\columnwidth]{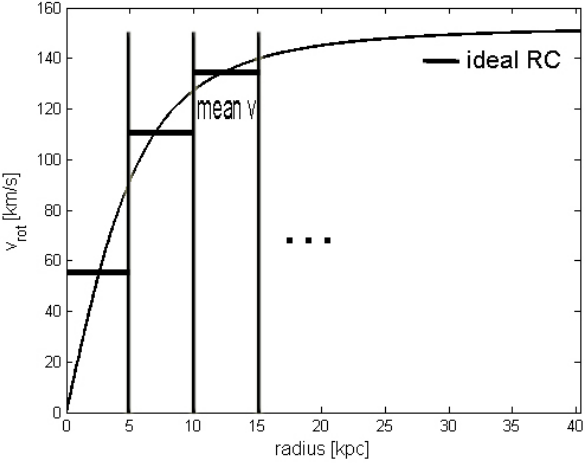}} \caption{An
example for a 5 kpc binning of the ideal RC.} \label{RCextraktor3}
\end{center}
\end{figure}

\noindent In order to determine systematic effects of large slit
widths d (relative to the galaxy size), which occur in the case of
observations of distant (z$\approx$0.5) galaxies (e.g. B\"ohm et
al. 2004) we vary d in a range of several kpc. Not only the
relative slit width varies in observations of distant disk
galaxies but also the sampling of the velocity field along the
spatial axis (i.e. the spatial resolution). In order to simulate
this finite spatial resolution we bin the reference RC with
different bin sizes. In Fig. \ref{RCextraktor3} we show the
extraction of different RCs, corresponding to different spatial
resolutions along the slit.

\section{Results}

\subsection{Rotation curves as a function of evolution}

\begin{figure}
\begin{center}
{\includegraphics[width=\columnwidth]{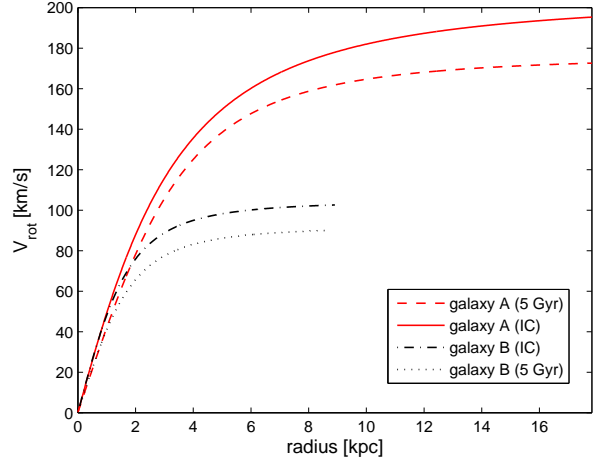}} \caption{Best
fits to the measured RCs of model galaxies A and B. The fits are
done for the ICs and for the evolved (5 Gyr) systems.}
\label{rot_evo}
\end{center}
\end{figure}

\begin{figure}
\begin{center}
{\includegraphics[width=\columnwidth]{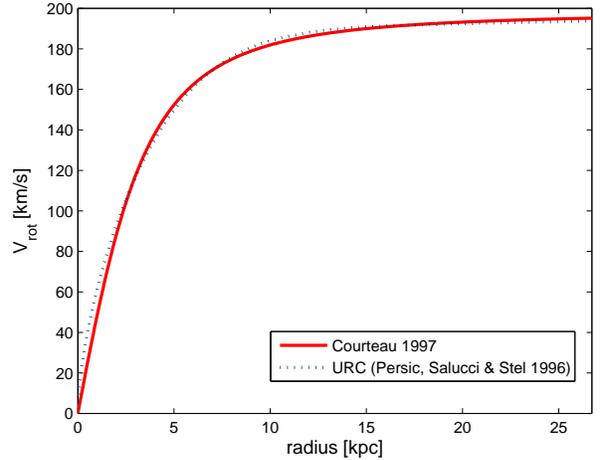}}
\caption{Comparison of the two profiles (Courteau, URC) for model
galaxy A. Both profiles represent the velocity field very well.}
\label{compare_fits}
\end{center}
\end{figure}

As a first step we investigate the RCs of the ICs and compare them
to the RCs of the fully hydrodynamically treated galaxies after 5
Gyr of evolution. Note that the ICs are based on an analytic model
introduced by Mo et al. (1998). The evolution starting from these
ICs is determined by the influence of the dynamics, the star
formation with feedback and stellar winds of the system. Therefore
the RCs and other internal properties of the evolved galaxies do
differ from those of the ICs. The general evolutionary trend of
the RCs is presented in Fig. \ref{rot_evo}. The rotation curves
were obtained by setting a slit width of 4 kpc (galaxy A) and 1
kpc (galaxy B) without any slit misalignment $\delta$. The
galaxies were inclined with an inclination angle of
i=$80^{\circ}$. A spatial resolution of 0.1 kpc was adopted to
extract ideal RCs. It is clearly visible, that the rotational
velocities get lower for the evolved galaxies. The decrease of the
overall angular momentum of gas particles in the disk can be
explained by mass ejection due to galactic winds and the
rearrangement of the gas in the disk due to the fully hydrodynamic
treatment as the disk rotates. As we want to investigate the
dependencies on different observational constraints, e.g. galaxy
alignment with respect to the spectroscopic slit or slit
misalignment, we use hereafter models to describe the shape of the
RC. Used in observational work are physically motivated fitting
functions, like the universal rotation curve (URC) eq. 2 (Persic
et al. 1996), or purely phenomenological fitting formulae like eq.
1 (Courteau 1997). Although we are aware that eq. 1 cannot
reproduce the many observed declining RCs, it is suitable for our
model galaxies, which do not decline in the outer parts. The RCs
shown in Fig. \ref{rot_evo} are best fits to the measured
rotational velocities of the model galaxies using the fitting
formula of Courteau (1997). This function is defined as

\begin{equation}
\rm{V}_{\rm{rot}}(r)=\frac{\rm{V}_{\rm{max}}r}{(r^a+r_{0}^a)^{\frac{1}{a}}}
\;\;\;\rm{[km/s]},
\end{equation}

\noindent where r is the galactocentric distance and $r_{0}$ and
$a$ are free fitting parameters. Clearly, with the general
decrease of the rotational velocities also V$_{\rm{max}}$ gets
lower. The physically motivated URC (Persic et al. 1996) is a
superposition of the velocity field of the DM halo and the disk.
The URC can be expressed as follows

\begin{eqnarray}
V_{\rm{URC}} \left( \frac{r}{r_{\rm{opt}}} \right) & = &
V(r_{\rm{opt}}) \left\{ \left( 0.72 + 0.44 \log(\frac{L}{L_{*}})
\right) \right. \nonumber \\
 & & \frac{1.97 x^{1.22}}{(x^2+0.78^2)^{1.43}} + 1.6 \exp[-0.4 \frac{L}{L_{*}}] \nonumber \\
 & & \left. \frac{x^2}{x^2+1.5^2 \left( \frac{L}{L_{*}} \right)^{0.4}}
 \right\}^{\frac{1}{2}} \;\;\;\rm{[km/s]},
\end{eqnarray}

\noindent where $x=r/r_{\rm{opt}}$, $V_{\rm{opt}}$ is the
rotational velocity at $r_{\rm{opt}}$ (the reference scale, which
for an exponential disc is 3.2$R_{\rm{d}}$) and $L$ the absolute
blue luminosity of the galaxy. In Fig. \ref{compare_fits} the URC
fit to our data is shown together with the Courteau (1997) fit.
Both functions represent the RCs of the model galaxies very well.
Although we are aware that the Courteau fitting function has no
physical justification, we use both functions to fit our data, in
order to investigate the shape of the RC as a function of the
observational setup, i.e. galaxy alignment and slit properties.

\noindent The fit parameters for eq. 1 and 2 with 95\% confidence
level errors are listed in Table \ref{fitting parameters} and 3,
respectively. Note that no restrictions were set on the fit
parameters (V$_{\rm{max}}$, r$_{0}$ and a for the Courteau
function and $V_{\rm{opt}}$ for the URC). The blue band luminosity
$L_B$ was estimated from the stellar mass assuming a stellar mass
to light ratio of 1.2 (mean from different star formation
histories in the redshift range $z=0.5-1.4$, Dickinson et al.
2003). The $L_*$ luminosity at $z=0.5$ was adopted as
$M_B^*=-21.3$ (Gabasch et al. 2004) this corresponds to
$\log(L_B^*)=10.72$. We find values for $L/L_*$ on the range of
0.61-0.65 for model galaxy A and 0.075-0.08 for model galaxy B. An
evolution of $L/L_*$, due to new forming stars in the galaxies,
can be seen. For the investigations of the RC as a function of
observational bias, we choose $L/L_*$ as an additional free
parameter, to get better
representations of the RCs.\\
Most present models of galaxy formation and evolution rely on the
work of Mo et al. (1998). Indeed, this model can reproduce
correctly the general shape of rotation curves, but galaxy
evolution can alter the RC. Note especially the differences
between V$_{\rm{c}}$ and V$_{\rm{max/opt}}$ from tables
\ref{galaxyproperties}, \ref{fitting parameters} and 3,
respectively. A comparison of semi-analytic models and
observations is generally complicated by the fact that
V$_{\rm{max/opt}}$ is determined differently from V$_{\rm{c}}$. In
the case of mass reconstruction via RCs the superposition of the
velocity field of the halo and the disk in the URC ansatz would be
the adequate approach.

\begin{table}
\caption[]{Fitting parameters for eq. 1 (Courteau 1997) for RCs of
the ICs and the fully hydrodynamically treated galaxies (evolution
5 Gyr).}
\begin{tabular}{c c c c }
\hline \hline Fitting & Parameter &  Galaxy A & Galaxy B \cr
\hline I & V$_{\rm{max}}$ [km/s] & 205$_{-3}^{+3}$ &
105$_{-0.8}^{+1.8}$ \cr I & r$_{0}$ [kpc] & 3.91$_{-0.14}^{+0.14}$
& 1.83$_{-0.08}^{+0.07}$ \cr I & a & 1.62$_{-0.11}^{+0.11}$ &
1.89$_{-0.15}^{+0.51}$\cr \hline\hline II & V$_{\rm{max}}$ [km/s]
& 169.67$_{-4.57}^{+4.53}$ & 91.8$_{-2.36}^{+2.2}$ \cr II &
r$_{0}$ [kpc] & 4.6$_{-0.47}^{+0.4}$ & 2.05$_{-0.25}^{+0.17}$ \cr
II & a & 3.39$_{-1.23}^{+1.24}$ & 2.14$_{-0.4}^{+0.41}$ \cr
\hline\hline
\end{tabular}
\newline I ... initial
conditions \newline II ... evolved galaxies (t=5 Gyr)
\label{fitting parameters}
\end{table}

\begin{table}
\caption[]{Parameter $V_{opt}$ for eq. 2 (Persic et al. 1996) for
RCs of the ICs and the fully hydrodynamically treated galaxies
(evolution 5 Gyr).}
\begin{tabular}{c c c c }
\hline \hline Fitting & Parameter &  Galaxy A & Galaxy B \cr
\hline I & V$_{\rm{opt}}$ [km/s] & 186.5$_{-0.8}^{+0.8}$ &
80.3$_{-2.3}^{+2.3}$ \cr I & L/L$_{*}$ & 0.61 & 0.075 \cr
\hline\hline II & V$_{\rm{opt}}$ [km/s] & 183.2$_{-6}^{+6}$ &
91.1$_{-2.2}^{+2.2}$ \cr II & L/L$_{*}$ & 0.65 & 0.08 \cr
\hline\hline
\end{tabular}
\newline I ... initial
conditions \newline II ... evolved galaxies (t=5 Gyr)
\label{fitting parameters_salucci}
\end{table}

\subsection{Rotation curves as a function of the slit width for model galaxy A}

\begin{figure}
\begin{center}
{\includegraphics[width=\columnwidth]{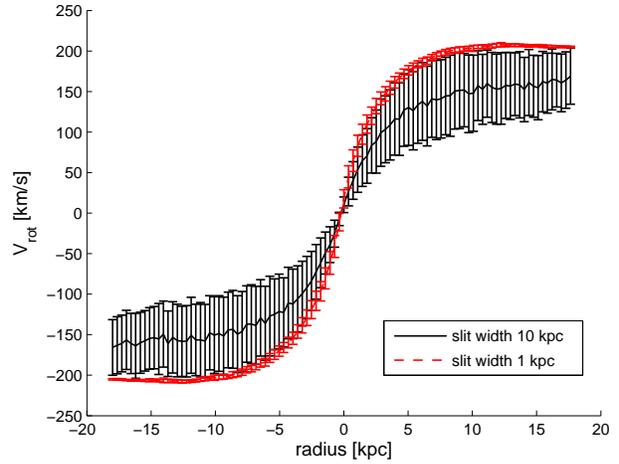}}
\caption{RCs of model galaxy A ICs, for two different slit widths
(1 kpc and 10 kpc). The error bars indicated in the figure are the
standard deviations of the mean velocity in each 0.1 kpc bin, see
Fig. \ref{RCextraktor2} (image b).} \label{rot_slit_1}
\end{center}
\end{figure}

\begin{figure}
\begin{center}
{\includegraphics[width=\columnwidth]{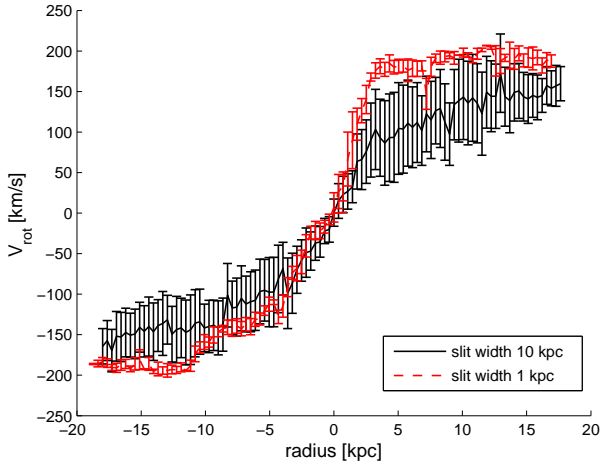}}
\caption{RCs of model galaxy A after 5 Gyr of evolution, for two
different slit widths (1 kpc and 10 kpc). The error bars indicated
in the figure are the standard deviations of the mean velocity in
each 0.1 kpc bin, see Fig. \ref{RCextraktor2} (image b).}
\label{rot_slit_2}
\end{center}
\end{figure}

In order to study systematic effects of large relative slit
widths, as they appear in observations of galaxies at intermediate
and high redshift, we extract the RC of model galaxy A for several
slit widths. Of course slit widths for local spiral systems are
orders of magnitudes smaller. But for observations of galaxies in
the redshift range 0.5 to 1, as carried out in a project related
to the present work (Ziegler et al., 2003), typical slit widths
become comparable to the disk scale length R$_{\rm{d}}$. It is
important to note, that such large slit widths result in an
integration of the velocity field perpendicular to the spatial
axis (slit direction). This effect is the optical equivalent to
'beam smearing' in radio observations. To investigate this effect we measure RCs
for slit widths ranging from 1 kpc to 10 kpc, which corresponds
for our model galaxy A to 0.2$h^{-1}$ R$_{\rm{d}}$ to 2.2$h^{-1}$
R$_{\rm{d}}$. In Fig. \ref{rot_slit_1} and \ref{rot_slit_2} we
show the RCs for model galaxy A (IC and evolved, respectively) for
two very different slit widths (1 kpc and 10 kpc).

\noindent If the slit width is 1 kpc the scatter around the mean
velocity in each bin is very small in comparison to the 10 kpc
slit. This can be explained in terms of velocity distributions in
a bin. As the slit width is increasing more gas particles can
contribute to the measured mean velocity in a bin. In other words,
the mean velocity is a superposition of velocity components from
different regions of the galaxy, mainly due to the line of sight
velocity distribution. In Fig. \ref{rot_slit_2} the same quantity
is shown as in Fig. \ref{rot_slit_1}, but after 5 Gyr of evolution
of model galaxy A. Again the same behaviour in the scatter and
mean velocity is present, but the overall velocity field shows
more structure. This is a consequence of the fully hydrodynamic
treatment of the galaxy. Note that ICs (as analytic models) do not
include prescriptions for spiral arms, which are present in
observed galaxies. Only the N-body/SPH simulations can reproduce
this feature. Therefore the measured ideal RC for the evolved
galaxies shows local fluctuations, connected to e.g. spiral arms.
This fact is well known from observations, where fluctuations of a
few tens of km/s are superposed on the smooth rotation curve of
the galaxy due to spiral arms (see e.g. Sofue and Rubin, 2001).
This is in good agreement with our model RCs (cf. Fig.
\ref{rot_slit_1} and Fig \ref{rot_slit_2}).

\begin{figure}
\begin{center}
{\includegraphics[width=\columnwidth]{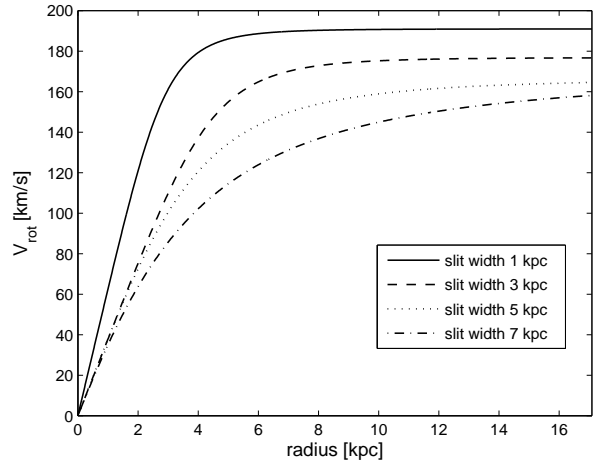}}
\caption{Best fits to RCs for different slit widths. The
underlying galaxy model is A after 5 Gyr of evolution.}
\label{rot_slit_3}
\end{center}
\end{figure}

\begin{figure}
\begin{center}
{\includegraphics[width=\columnwidth]{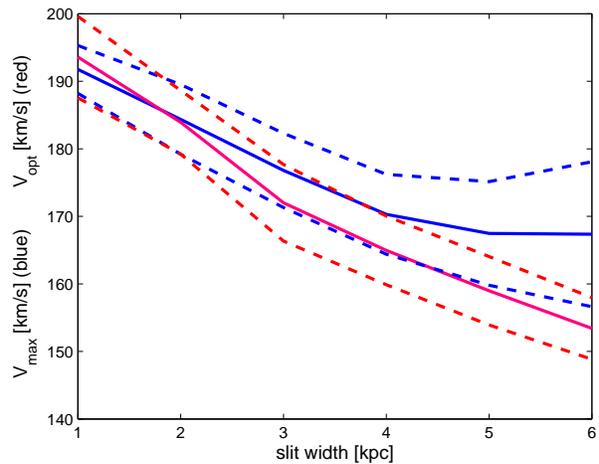}}
\caption{Fitting parameters V$_{\rm{max}}$ (blue) and
$V_{\rm{opt}}$ (red) as a function of slit width for model galaxy
A with ideal spatial resolution. The dashed lines correspond to
the bounds of the 95\% confidence level. After an almost linear
decrease the scatter increases for V$_{\rm{max}}$.}
\label{rot_slit_4}
\end{center}
\end{figure}

As a next step we fit eq. 1 to the RCs, extracted from different
slit widths, shown in Fig. \ref{rot_slit_3}. We used galaxy model
A after 5 Gyr of evolution. The galaxy was always 'observed' with
an inclination $i=80^{\circ}$, with an ideal spatial resolution of
0.1 kpc. As the slit width increases the fitted curves decrease.
Again this can be explained by the averaging process. It is
obvious that too wide slits (d$>$R$_{\rm{d}}$) result in non-flat
RCs, which should not be fitted by eq. 1. Instead an observer
would use here the URC. If we adopt our fitting procedures and
extract V$_{\rm{max}}$ and V$_{\rm{opt}}$ for different slit
widths, we obtain a dependence of V$_{\rm{max}}$ and
V$_{\rm{opt}}$ on d, as shown in Fig. \ref{rot_slit_4}. A nearly
linear decrease of V$_{\rm{max}}$ and V$_{\rm{opt}}$ from d=1 kpc
to d=4.5 kpc can be seen. From this investigation we would
recommend to apply only a maximum slit width in the order of
R$_{\rm{d}}$.

\subsection{Rotation curves as a function of inclination}

\begin{figure}
\begin{center}
{\includegraphics[width=\columnwidth]{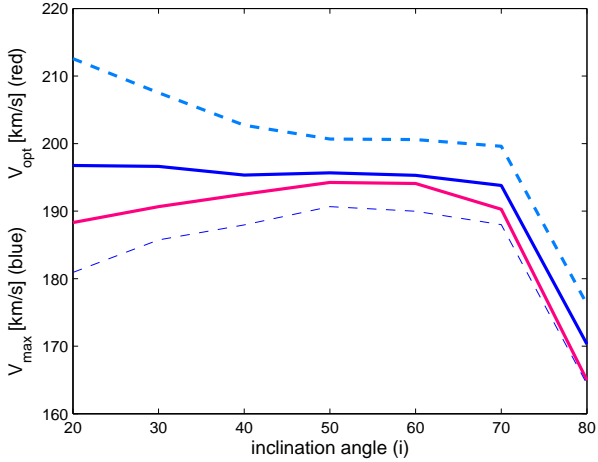}}
\caption{Fitting parameter V$_{\rm{max}}$ (blue) and
$V_{\rm{opt}}$ (red) as a function of the inclination angle. The
dashed lines correspond to the bounds of the 95\% confidence level
of $V_{\rm{max}}$. The underlying model is galaxy A, after 5 Gyr
of evolution. A constant slit width of 4 kpc was used.}
\label{incl_1}
\end{center}
\end{figure}

\begin{figure}
\begin{center}
{\includegraphics[width=\columnwidth]{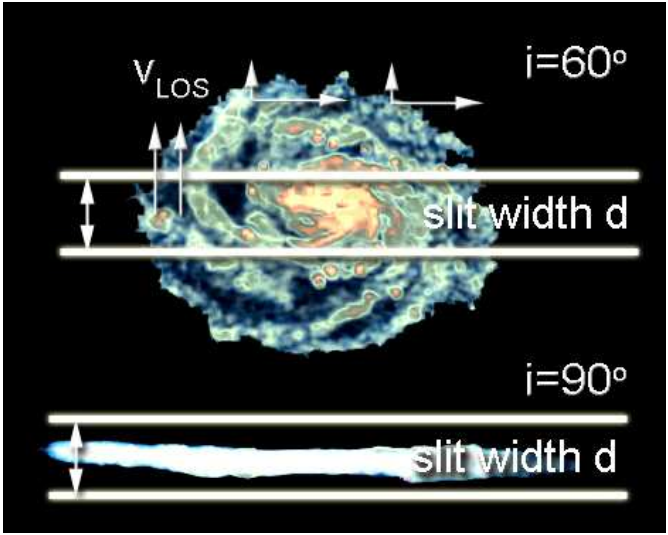}}
\caption{Sketch of the influence of the inclination on a spiral
galaxy for a fixed slit width for model galaxy A.} \label{incl_2}
\end{center}
\end{figure}

Galaxies are very rarely observable edge-on and therefore a
correction of inclination effects on the RC is important. The
intrinsic RC V$^{\rm{int}}$(r) of a galaxy is most often corrected
by the sine of the inclination angle, i.e. by the simple geometric
correction V$^{\rm{int}}$(r)=V$^{\rm{obs}}$(r)/sin(i) (edge-on
galaxies are defined to have i=$90^{\circ}$). We investigate the
RC of model galaxy A by rotating the galaxy from i=$20^{\circ}$ to
i=$80^{\circ}$, which is a typical range of inclination angles
accessible in observations. The slit was centred on the gas disk
with a slit width fixed to d=4 kpc. After each rotation we extract
the RC, corrected with the simple expression
V$^{\rm{int}}$(r)=V$^{\rm{obs}}$(r)/sin(i) before, and fit the
data with eq. 1 and 2. The dependence of V$_{\rm{max}}$ and
V$_{\rm{opt}}$ on the inclination angle $i$ is shown in Fig.
\ref{incl_1}. In the inclination range 80$^{\circ}>i>70^{\circ}$
V$_{\rm{max}}$  and V$_{\rm{opt}}$ show a steep increase while for
$i$ $<$ 65$^{\circ}$ V$_{\rm{max}}$ and V$_{\rm{opt}}$ stay
roughly constant. The explanation for this behavior is shown in
Fig. \ref{incl_2}. If a galaxy is observed nearly edge-on, the
rotational velocity is an average of velocities from all radial
distances along the line-of-sight. With decreasing inclination,
more and more gas particles with lower line-of-sight velocity
components move out of the slit, and are therefore not taken into
account in the averaging process. Thus, the mean velocity in each
bin increases, which again leads to a larger V$_{\rm{max}}$ and
V$_{\rm{opt}}$. Below a certain inclination angle, depending on
the slit width, most of the volume of the disk is not covered by
the slit. In this volume most gas particles with a low
line-of-sight velocity component are located. In our case 70\% of
the volume is not covered by the slit for $i<65^{\circ}$. Note
that this behavior is most significant for large relative slit
widths. In observations one tries to overcome the problem
occurring at high inclination angles by special techniques, as
e.g. the 'envelope-tracing' method (e.g. Sofue and Rubin, 2001).

\begin{figure}
\begin{center}
{\includegraphics[width=\columnwidth]{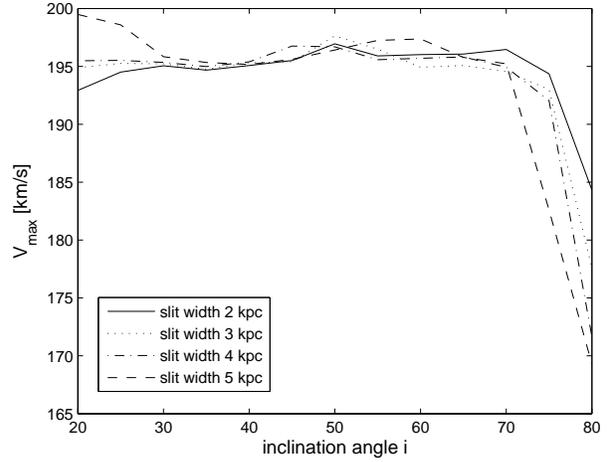}}
\caption{Fitting parameter V$_{\rm{max}}$ as a function of the
inclination angle, for slit widths in the range of 2$\leq$d$\leq$5
kpc for model galaxy A.} \label{incl_3}
\end{center}
\end{figure}

\noindent As the inclination angle $i$ decreases the errors for
the fit become larger. Note that for $i<35^{\circ}$ the errors are
in the order of 10\%. In Fig. \ref{incl_3} we show the fitting
parameter V$_{\rm{max}}$ as a function of the inclination angle
$i$ for different slit widths. If the inclination is below
70$^{\circ}$ and above 30$^{\circ}$ the slit width does not affect
V$_{\rm{max}}$. The same behaviour was found for V$_{\rm{opt}}$.
Only in the cases near edge on and face on, the slit width plays
an important role in covering gas particles. The overall trend is
the same as shown in Fig. \ref{rot_slit_4}, where larger slit
widths lead to lower V$_{\rm{max}}$ and larger errors.

\noindent Therefore, for large inclinations ($i > 70^{\circ}$ the
slit width varies V$_{\rm{max}}$  strongly (see Fig. \ref{incl_2}
and \ref{incl_3}), while for smaller inclinations the
determination of V$_{\rm{max}}$ does not strongly depend on the
slit width.

\subsection{Rotation curves as a function of misalignment}

\begin{figure}
\begin{center}
{\includegraphics[width=\columnwidth]{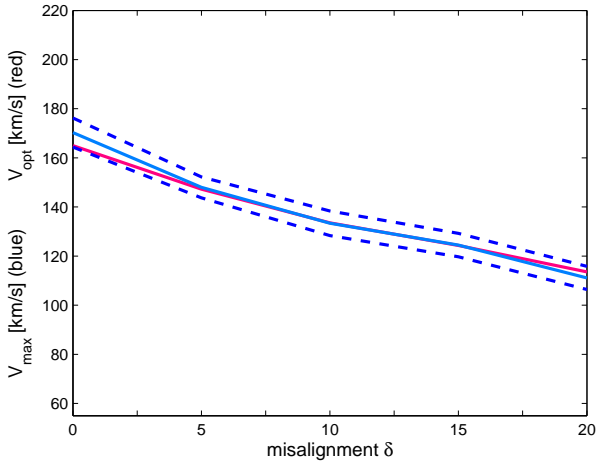}} \caption{
Fitting parameters V$_{\rm{max}}$ (blue) and V$_{\rm{opt}}$ (red)
as a function of the misalignment angle, for a fixed slit width of
d=4 kpc for model galaxy A. The dashed lines correspond to the
bounds of the 95\% confidence level for the fitting parameter
$V_{\rm{max}}$.} \label{slit_mis}
\end{center}
\end{figure}

As shown in Fig. \ref{RCextraktor1} the slit for measuring the RCs
can have a misalignment $\delta$, with respect to the major axis
of the projection of the galaxy on the sky. Especially
multi-object spectroscopy has to deal with misaligned slits,
therefore we investigate the effect of $\delta$ on the fitting
parameters V$_{\rm{max}}$ and $V_{\rm{opt}}$. In Fig.
\ref{slit_mis} we plot the result for different $\delta$ for a
fixed slit width of d=4 kpc for model galaxy A. Note that we
applied the standard $\cos^{-1}(\delta)$ correction. Nevertheless
the fitting parameters V$_{\rm{max}}$ and $V_{\rm{opt}}$ are not
independent of $\delta$. The corresponding error (scatter around
the mean velocity in a bin), does not show any dependence on the
misalignment angle $\delta$, if it covers particles all along the
slit. As the $cos(\delta)$ correction, that we have applied, is
only fully valid for two dimensional disks, without any thickness,
V$_{\rm{max}}$ and $V_{\rm{opt}}$ show a dependence on $\delta$,
the slit misalignment. In the case of multi-object spectroscopy,
where the misalignment can be much higher, more advanced
corrections have to be applied. We have introduced one method in
B\"ohm et al. (2004).

\subsection{Rotation curves as a function of binning}

\begin{figure}
\begin{center}
{\includegraphics[width=\columnwidth]{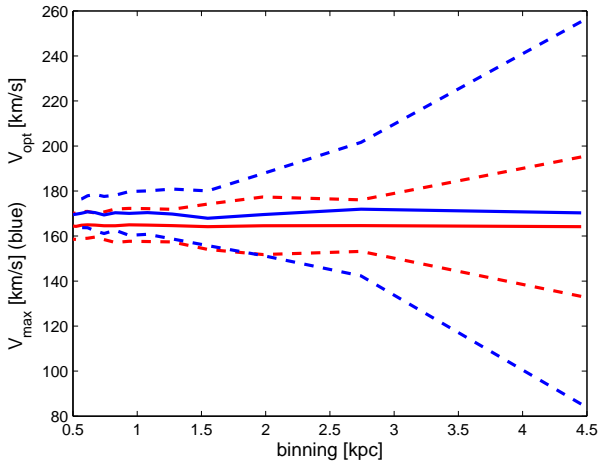}}
\caption{Fitting parameters V$_{\rm{max}}$ and V$_{\rm{opt}}$ as a
function of the spatial resolution. Obviously the quality of an RC
decreases for poorer spatial resolutions.} \label{binning}
\end{center}
\end{figure}

To simulate different spatial resolution we bin our ideal RC, see
Fig. \ref{RCextraktor3}. In Fig. \ref{binning} we show the
dependency of the fitting parameters $V_{\rm{max/opt}}$ on
different spatial resolutions, together with standard deviations.
Obviously a poor spatial resolution leads to larger errors, but
the value V$_{\rm{max}}$ shows no dependence. As V$_{\rm{max}}$
represents the flat part of the RC (the asymptotic velocity), the
binning does not vary V$_{\rm{max}}$, but of course a and r$_{0}$.

\subsection{Estimators for the virial mass}\label{virmass}

One major goal of investigating RCs is the possibility to
determine the mass of the system including baryonic and
non-baryonic components. The virial mass of a galaxy is $\propto$
to V$_{c}^3$ (e.g. Mo et al. 1998), where V$_{c}$ is the
rotational velocity of the galaxy at the virial radius.
Unfortunately, it is not possible to measure this velocity
directly, therefore the virial mass of a galaxy is estimated by
some suitable measure of the maximum circular velocity. If
V$_{\rm{max}}$ converges to V$_{\rm{c}}$ at the virial radius, the
RC fitting procedure with eq. 1 is a good estimator for the virial
mass. However, as V$_{\rm{max}}$ depends on observational
constraints and of internal properties of the galaxy, it seems to
be no robust estimator for the total virial mass. Van den Bosch
(2002) investigated the impact of cooling and feedback on disk
galaxies using an analytical model. The author comes to similar
conclusions, however, as he does not treat the gas self
consistently (i.e. fully hydrodynamically), he cannot derive a
rotation curve and V$_{\rm{max}}$ as we do. Instead in van den
Bosch's work V$_{\rm{max}}$ is defined as the maximum rotation
velocity inside the radial extend probed by the cold gas (van den
Bosch, 2002). The author finally concludes that eq. 3 seems to be
a good fitting function to obtain the virial mass of a system.

\begin{equation}
\rm{M}_{\rm{vir}}=2.54\times 10^{10} \rm{M_{\sun}}
\left(\frac{r_d}{kpc}\right) \left( \frac{V_{max}}{100 km s^{-1}}
\right)^2
\end{equation}

\noindent Van den Bosch (2002) did not intend to give an exact
estimator of the virial mass by this fitting function. Instead, it
seems that he wanted to give a rough estimate for the influence of
cooling and feedback on the determination of the virial mass using
V$_{\rm{max}}$. Therefore, it is clear that eq. 3 is not exact.
Nevertheless, as some observers use this function we also
apply it to our simulated galaxies.\\
As we treat the galaxies fully hydrodynamically, we are able to
show the influence of evolution on the RCs. In Fig. \ref{rot_evo}
the general decrease of V$_{\rm{max}}$ is given. The main part,
which decreases the angular momentum is the presence of a galactic
wind, which ejects matter from the disk into the surrounding halo.
In most cases we find that V$_{\rm{max}}$ derived from the fitting
procedures does not represent V$_{\rm{c}}$ and is also different
from the maximum rotation velocity present in the full velocity
field. Most determinations of V$_{\rm{max}}$ (different
inclination, slit misalignments and slit widths) result in an
overestimation of V$_{\rm{c}}$, if the standard corrections
$\sin(i)$ and $\cos(\delta)$ are applied. The virial mass of our
model galaxies, obtained by extracting r$_{200}$ as limiting
radius, yields to M$_{200}$=1.13$\times$10$^{12}$ M$_{\sun}$ for
model galaxy A and M$_{200}$=1.44$\times$10$^{11}$ M$_{\sun}$ for
model galaxy B. We used eq. 2 to determine the virial mass and
find a systematic underestimation of $\approx$ 50\%. As van den
Bosch (2002) states correctly, matter can be ejected by galactic
winds and therefore reduce V$_{\rm{max}}$ by increasing at the
same time R$_{\rm{d}}$. The same behavior is present in our
N-body/SPH simulations. However, we find that eq. 3 underestimates
the virial mass of our model galaxies. The same underestimation of
van den Bosch's (2002) estimation was stated by Conselice et al.
(2005). However, as we have mentioned earlier, eq. 3 is not
thought to be exact. In fact, van den Bosch (2002) mentions that
the error for an individual galaxy can still exceed a factor of 2.
Thus it is no surprise, that the result is not correct for our
model galaxy.\\
A more detailed mass decomposition by applying the URC fitting
would allow a deeper insight into the mass distribution of a
galaxy. Nevertheless, the same problems from the observational
point of view would be inherent, like slit width, slit
misalignment or galaxy orientation. A detailed investigation of
the influence of the previous mentioned effects on the mass
decomposition by the RC will be investigated. Here we conclude
that it is important for observers to investigate the environment
of the measured galaxy. As galactic winds can strongly influence
the internal kinematics of the gas in the disk, the knowledge of
the star formation rate gives constraints on the robustness of the
determination of the virial mass from V$_{\rm{max}}$. An important
issue in this context is the membership of the galaxy to a group,
a galaxy cluster or the field. It is important to note that
especially spiral galaxies in galaxy clusters can often interact
with each other. As the interaction (Kapferer et al. 2005)
increases the star formation rate significantly, merger-driven
starbursts occur for a short (up to several 100 million years)
time and expel huge amounts of inter stellar matter into the
surrounding halo. Therefore, these systems might have lower
rotational velocities than isolated, low star forming galaxies
while the general shape of the RC remains similar.

\subsection{The Tully-Fisher relation of spiral galaxies}\label{TF}

\begin{figure}
\begin{center}
{\includegraphics[width=\columnwidth]{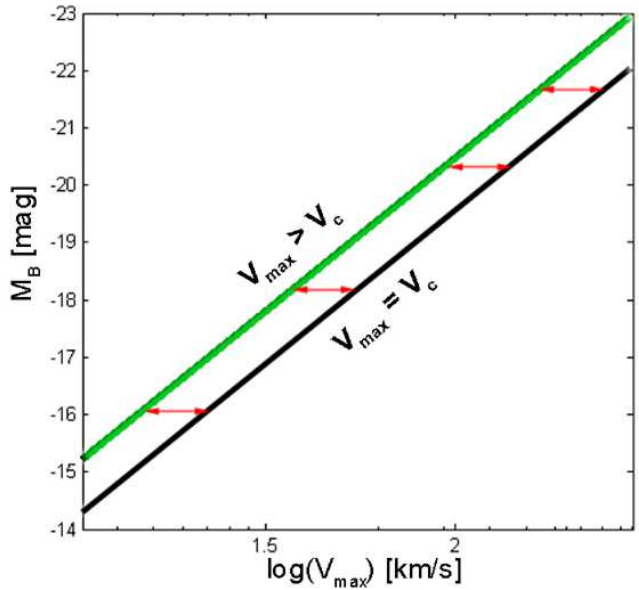}} \caption{Sketch
of a shift in the Tully-Fisher relation due to discrepancies
between V$_{\rm{c}}$ and V$_{\rm{max}}$. Only if the assumption
V$_{\rm{max}}$=V$_{\rm{c}}$ holds, Tully-Fisher relations from
semi-analytical models and from observations are comparable. If
V$_{\rm{max}}$$>$V$_{\rm{c}}$ (result of our investigation) the
line of the semi-analytical model would be shifted to the left
with respect to the observed line.} \label{tully}
\end{center}
\end{figure}

Semi-analytical galaxy formation models usually fix their free
parameters such that the models match observed present-day
luminosity functions (LF) or TFRs. Early models had problems
predicting at the same time the correct LF and the correct zero
point of the TFR. Therefore, additional physical processes were
introduced to fit both important statistical properties of a
galaxy population at the same time. However, most semi-analytical
models only have rough models to approximate the circular velocity
of the halo V$_{\rm{c}}$, to derive a TFR for the underlying
galaxy population. This in turn, can differ significantly from the
maximum rotational velocity V$_{\rm{max}}$, derived from observed
RCs, as discussed in the previous sections. Thus, differences of
the zero point of observed and modelled TFRs could, at least to
some extend, origin from these discrepancies of V$_{\rm{max}}$ and
V$_{\rm{c}}$. We do not claim here, that the differences between
V$_{\rm{max}}$ and V$_{\rm{c}}$ solve the problem of the zero
point discrepancies of observed and simulated TFRs. We just point
out that these two quantities are in general not equal and
therefore it is problematic to use them equivalently. A major
contribution to the difference between V$_{\rm{max}}$ and
V$_{\rm{c}}$ are galactic winds, as they are able to decrease
angular momentum of a disk, by expelling a significant amount of
matter into the surrounding halo. As we are now able to show, that
for given modelled galaxies, V$_{\rm{max}}$ as derived in
observations, indeed mostly overestimates V$_{\rm{c}}$, we
emphasise the importance of cooling and feedback processes. We
plan to investigate the TFR of a semi-analytical model, taking
this discrepancy into account, in a forthcoming work. The effect
of the overestimation of V$_{\rm{max}}$ as V$_{\rm{c}}$ is
sketched in Fig. \ref{tully}. If V$_{\rm{max}}$ was equal to
V$_{\rm{c}}$ we would expect that observed and simulated TFRs
coincide in the right line. If V$_{\rm{max}}$ is higher than
V$_{\rm{c}}$ the TFR of the simulations would be located at the
position of the left indicated line.

It is important to stress, that the direct comparison of
observation (V$_{\rm{max}}$) and simulation (V$_{\rm{c}}$) does
invoke uncertainties originating from internal kinematics of the
disk, like galactic winds. Similar concerns were emphasized by van
den Bosch (2000). Tully Fisher relations can also be constructed
by using $V_{\rm{opt}}$. Again the same observational constraints
would introduce uncertainties, but the increase of the disk scale
length due to galactic winds is compensated by measuring
$V_{\rm{opt}}$ at a given optical radius. Different disk scale
lengths result in different optical radii and therefore the
evolution of the gaseous disk is taken into account.

\section{Summary and conclusion}

In this work we use fully hydrodynamically modelled galaxies,
including star formation, stellar feedback and galactic winds to
study the internal kinematics of the gas in a spiral galaxy. We
extract an RC from the line-of-sight velocities of the gas
particles. Then we use a three-parameter fitting formula to
describe the rotation curve (Courteau, 1997) and the universal
rotation curve (URC, Persic et al. 1996). We find

\begin{itemize}

\item{that for evolved model galaxies the extracted and fitted RCs
show a tendency to lower rotational velocities, compared to the
initial conditions. This can be explained by galactic winds, which
expel a certain amount of matter into the surrounding halo and
therefore decrease the total angular momentum.}

\item{We show that the variation of the slit width does influence
the quality of the RC and the values of V$_{\rm{max}}$ as well as
$V_{\rm{opt}}$. If the slit width is small, V$_{\rm{max}}$ and
$V_{\rm{opt}}$ are systematically higher and show less scatter. If
the slit width is large, the mean velocity is a superposition of
particles along the line of sight, which introduces low velocity
components and therefore larger scatter.}

\item{The dependence of V$_{\rm{max}}$ and $V_{\rm{opt}}$ on the
inclination angle is nearly constant over the range of
$20^{\circ}<i<70^{\circ}$. In the range above $i=70^{\circ}$ both
show a strong dependence on $i$, which can be explained by
rotating velocity components from the foreground and background of
the disk into the slit. It turned out that the sine correction
leads to similar results in the range of
$20^{\circ}<i<70^{\circ}$. In this range we do not encounter
strong dependencies on the slit width.}

\item{The $cos(\delta)$ correction for the slit misalignment is
only fully valid for two dimensional disks, without any thickness.
Thus, V$_{\rm{max}}$ and $V_{\rm{opt}}$ show an almost linear
dependence on $\delta$. In the case of multi-object spectroscopy,
where misalignments are inherent, more advanced corrections have
to be applied, as for example introduced by B\"ohm et al. (2004).}

\item{The spatial resolution does not influence V$_{\rm{max}}$ and
$V_{\rm{opt}}$ strongly, but influences the quality of the RC.}

\item{We test the capability of V$_{\rm{max}}$ as an estimator for
the virial mass of the system and found a strong overestimation of
the virial mass, by applying the virial theorem. By testing a more
sophisticated relation, including results of semi-analytic models,
introduced by van den Bosch (2002), we find an underestimation in
the order of 50\% of the virial mass. The explanation for the
disagreement with van den Bosch lies in the fully
N-body/hydrodynamic treatment in our simulations. Another point is
the discrepancy of deriving V$_{\rm{max}}$, in our case from RCs
from our model galaxies and in his case of the semi-analytical
approach.}

\item{As Tully-Fisher relations are a common tool for testing
models of galaxy evolution, any systematic differences between
observations and theory play an important role. We show that
V$_{\rm{max}}$ and $V_{\rm{opt}}$ usually differ from
V$_{\rm{c}}$, which introduces a shift into the Tully-Fisher
relation.}

\end{itemize}

The investigation of the influence of minor/major mergers and
galaxy flybys is an important issue on this topic, which will be
discussed in a forthcoming paper.

\section*{Acknowledgements}

The authors would like to thank Volker Springel for providing them
GADGET2 and his initial conditions generators. generators. The
authors are grateful to the anonymous referee for his/her
criticism that helped to improve the paper. The authors
acknowledge the Austrian Science Foundation (FWF) through grant
number P15868, the UniInfrastrukturprogramm 2004 des bm:bwk
Forschungsprojekt Konsortium Hochleistungsrechnen, the bm:bwk
Austrian Grid (Grid Computing) Initiative and the Austrian Council
for Research and Technology Development and the German Science
Foundation (DFG) through Grant number Zi 663/6-1. In addition the
authors acknowledge the Deutsches Zentrum f\"ur Luft- und
Raumfahrt through grant 50 OR 0301, the ESO-Mobilit\"atsstipendien
des bm:bwk (Austria) and the Tiroler Wissenschaftsfonds.

\end{document}